\documentclass{cjpsuppl}
\usepackage{epsf}

\def\als{\alpha_s}

\def\gev{\,{\rm GeV}}
\def\vk{{\bf k}_{\perp}}

\begin{document}
\title{Spin effects in vector meson and  $Q \bar Q$ production}
\authori{S.\,V. Goloskokov}
\addressi{Bogoliubov Laboratory of Theoretical  Physics,
  Joint Institute for Nuclear Research, Dubna, Russia}
\authorii{}    \addressii{}
\authoriii{}   \addressiii{}
\authoriv{}    \addressiv{}
\authorv{}     \addressv{}
\authorvi{}    \addressvi{}
\headtitle{Spin effects in vector meson production and \ldots}
\headauthor{S.\,V. Goloskokov} \lastevenhead{S.\,V. Goloskokov:
Spin effects in vector meson and  $Q \bar Q$ production \ldots}
\refnum{}
\daterec{20 October 2002;\\final version 31 December 2003}
\suppl{A}  \year{2003} \setcounter{page}{1}
\maketitle

\begin{abstract}
We study light hadron leptoproduction at small $x$. Vector meson
production is analysed in terms of generalized gluon distributions
with taking into account the transverse quark motion. Within a
two-gluon model the double spin asymmetries for longitudinally
polarized leptons and transversely polarized protons in the
diffractive $Q \bar Q$ production are investigated. The predicted
$A_{lT}$ asymmetry is large and can be used to obtain information
on the polarized gluon distributions in the proton.
\end{abstract}

\section{Introduction}
In this report, we analyse diffractive hadron leptoproduction at
high energies and small $x$. Intensive experimental study of
diffractive processes was performed in DESY (see e.g.
\cite{zeus97,h1_99,zeus} and references therein). The amplitude of
vector meson production for longitudinally  polarized photon and
vector meson give a predominant contribution to the cross section
as $Q^2 \to \infty$. It can be expressed in terms of generalized
parton distribution (GPD) in the nucleon ${\cal F}_\zeta(x)$ where
$x$ is a fraction of the proton momentum carried by the outgoing
gluon and $\zeta$ is the difference between the gluon momenta
(skewedness) \cite{rad-j}. It was found within the GPD approach
\cite{mpw} and two-gluon model \cite{fra95} that the leading-twist
contribution to the cross section is far from experiment. The
cross sections of diffractive quark- antiquark production at small
$x$ \cite{die95, bart96, ryskin97} were found to be expressed in
terms of the same gluon distributions as for the vector meson
production. Such processes give a possibility to study properties
of GPD at small $x$. Note that in polarized reactions
spin-dependent parton distributions can be investigated.

We discussed our results for vector meson production at small $x$
within GPD approach presented in \cite{gol_kroll}. It is shown
that the inclusion of transverse degrees of freedom leads to
reasonable agreement of theory with experiment for light meson
production. The double spin asymmetries for longitudinally
polarized leptons and transversely polarized protons in
diffractive $Q \bar Q$ production at high energies are
investigated within the two-gluon exchange model. The predictions
are made for a future eRHIC collider with polarized lepton and
proton beams (see e.g. \cite{erhic}) where this asymmetry can be
studied.

\section{Hadron production at small $x$ and GPD}

Let us study the diffractive hadron production in lepton-proton
reactions
\begin{equation}
\label{react} l+p \to l+p +H
\end{equation}
at high energies. Here the hadron state $H$ can contain either a
vector meson or a $Q \bar Q$ system which are observed as two
final jets. The reaction (\ref{react}) can be described in terms
of the kinematic variables which are defined as follows:
\begin{eqnarray}
\label{momen} q^2= (l-l')^2=-Q^2,\;t=r_P^2=(p-p')^2, \nonumber
\\  y=\frac{p \cdot q}{l  \cdot p},\;x=\frac{Q^2}{2p  \cdot q},\;
x_P=\frac{q \cdot (p-p')}{q \cdot p},\; \beta=\frac{x}{x_P},
\end{eqnarray}
where $l, l'$ and $p, p'$ are the initial and final lepton and
proton momenta, respectively, $Q^2$ is the photon virtuality, and
$r_P$ is the momentum carried by the two-gluons (Pomeron). The
variable $\beta$ appears in the $Q \bar Q$ production. In this
case, the effective mass of a produced quark system
$M_X^2=(q+r_P)^2$ might be  large and  $\beta=x/x_P \sim
Q^2/(M_X^2+Q^2)$ can vary here from 0 to 1. For the diffractive
vector meson production, $M_X^2=M_V^2$ and $\beta \sim 1$ for
large $Q^2$. Here the momentum $V$ is on the mass shell
$V^2=(q+r_P)^2=M_V^2$ and $x_P$ determined by
\begin{equation}
\label{x_P} x_P \sim \frac{m_V^2+Q^2+|t|}{s y}
\end{equation}
is small at high energies. The $x_P$ is not fixed for the $Q \bar
Q$ production.

The  light hadron production at small $x$ is predominated by the
gluon GPD. The gluon contribution to the helicity amplitudes for
leptoproduction of longitudinally polarized vector mesons looks
like
\begin{eqnarray}\label{amp-nf}
{\cal M}^{V(g)}_{0+,0 +} &=& e\, \sum_a e_a B^V_a\,
        \sqrt{1-\zeta}\, \int_0^1 \frac{dX \;\;  {\cal H}^{(g)}_{0+,0 +}\,
                                   {\cal F}^g_\zeta(X,t)}{(X-i {\varepsilon})
                  (X-\zeta + i {\varepsilon})}.
\end{eqnarray}

The leading twist parton-level amplitudes have the form

\be {\cal H}^{(g)}_{0+,0 +}\, = \,\frac{4\pi \als(Q/2) f_V}{N_c Q}
           \,\int_0^1 d\tau\, \frac{\Phi_V(\tau)}{\tau(1-\tau)}
           \,,
\label{sub-amp} \ee where $\Phi_V$ is the distribution amplitude.

A model for the gluonic GPD is constructed by using a double
distribution \cite{rad99a}
 \be {\cal F}^g_\zeta(X,t) =
\int_0^1\,dx\,\int_0^{1-x}\,dy\, f^g(x,y,t)\,
\delta(X-x-\zeta\,y)\,, \label{f}\ee where the skewdness $\zeta
\sim x_P$. The double distribution function is determined as
 \be f^g(x,y,t) = 6
\frac{y(1-x-y)}{(1-x)^3}\, x g(x) \, f(t)\,. \label{f1} \ee Here a
factor $f(t)$ models the $t$-dependence, $x g(x)$ is an ordinary
gluon distribution which at small $x$, is approximated by
 \be \label{xgxq2}
x g(x,Q^2)\simeq 1.94 x^{-0.17-0.05 \ln(Q^2/Q_0^2)},\;\;
\mbox{with }\;Q_0^2=4\gev^2.\ee

We  consider the quark and antiquark transverse  momenta  in the
meson to  modify the leading-twist amplitude. The distribution
amplitude $\Phi_V$ is replaced  in this case by  a Gaussian
parameterization \cite{jak93} \be
          \Psi_V(\tau,\vk)\,\propto  a^2_V
       \, \exp{\left[-a^2_V\, \frac{\vk^{\,2}}{\tau(1-\tau)}\right]}\,.
\label{wave-l} \ee Transverse momentum integration of
(\ref{wave-l}) leads  the asymptotic form of a meson distribution
amplitude $\Phi_V^{AS}=6\tau\bar{\tau}$. For the decay constants
$f_V$ we take the values $0.216 {\rm GeV}$ and $0.237{\rm GeV} $
for $\rho$ and $\phi$ mesons, respectively. For the transverse
size parameter $a_\rho$ we use a value of $0.6 \,\gev^{-1}$ which
leads to an average transverse momentum of $0.6\gev$. For the
$\phi$- meson production we take
$a_\phi=0.45 \,\gev^{-1}$ \cite{gol_kroll}.\\[0.5cm]
\begin{figure}
\centering \mbox{\epsfysize=80mm\epsffile{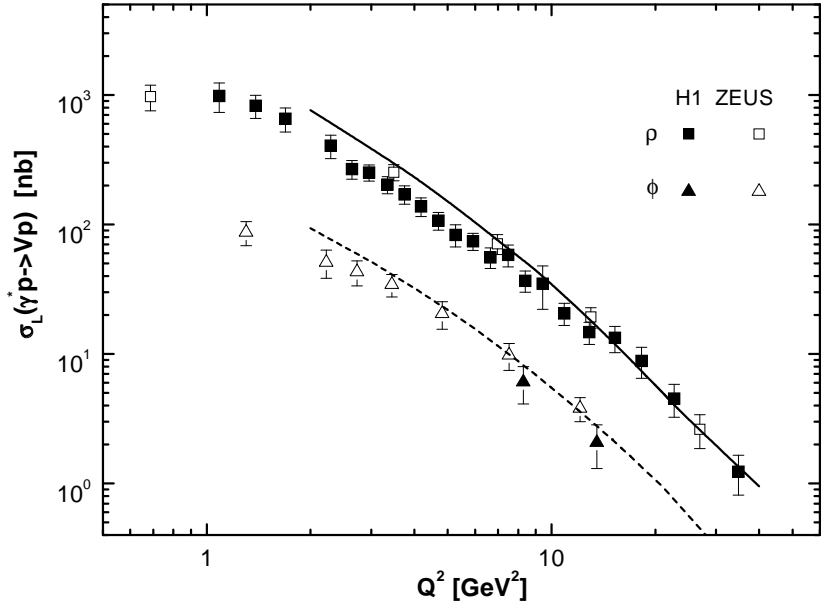}}
\caption{The longitudinal cross sections  at $W=75 {\rm GeV}$
extracted in \cite{zeus}.   The solid (dashed) line represent our
results for $\rho$ ($\phi$) production.} \label{kt_h}
\end{figure}

In the quark propagators of the subprocess  we  consider the
transverse momenta which are accompanied by the Sudakov factor.
The last factor suppresses contributions from the end-point
regions, in which one of the partons entering the meson wave
function becomes soft and  factorization breaks down.

The result of our calculations for the longitudinal cross sections
of the $\rho$ and $\phi$ production is shown in Fig. 1.  Good
agreement with experimental data  extracted from $\gamma^* p \to V
p$ in \cite{zeus} was obtained. Thus, the perturbative approach
with including a transverse quark momentum leads to agreement of
the theoretical results with the experiment.

\section{Diffractive $Q \bar Q$ production. Two-gluon exchange model}
Now let us study  the process of $Q \bar Q$ leptoproduction within
the two-gluon exchange model. As we have discussed previously,
this contribution is predominated at small $x \leq 0.1$ by a gluon
exchange. To study spin effects in diffractive hadron production,
one must know the structure of the two-gluon coupling with the
proton at small $x$. The two-gluon coupling with the proton which
describes transverse spin effects in the proton can be
parametrized in the form \cite{golostr}
\begin{eqnarray}\label{ver}
V_{pgg}^{\alpha\beta}(p,t,x_P,l_\perp)&=& B(t,x_P,l_\perp)
(\gamma^{\alpha} p^{\beta} + \gamma^{\beta} p^{\alpha}) \nonumber\\
&+&\frac{i K(t,x_P,l_\perp)}{2 m} (p^{\alpha} \sigma ^{\beta
\gamma} r_{\gamma} +p^{\beta} \sigma ^{\alpha \gamma}
r_{\gamma})+...  .
\end{eqnarray}
Here $m$ is the proton mass and $l$ is a gluon momentum. In the
matrix structure (\ref{ver}) we wrote only the terms with the
maximal powers of a large proton momentum $p$ which are symmetric
in the gluon indices $\alpha,\beta$. The structure proportional to
$B(t,...)$ determines the spin-non-flip contribution. The term
$\propto K(t,...)$ leads to the transverse spin-flip at the
vertex. The spin-dependent cross section of diffractive processes
is expressed in terms of the soft gluon coupling (\ref{ver}) which
is convoluted with the hard hadron production amplitude.

The spin-average and spin dependent cross sections with
longitudinal polarization of a lepton and different polarization
of a proton are determined by
\begin{equation}
\label{spm} \sigma(\pm) =\frac{1}{2} \left( \sigma(^{\rightarrow}
_{\Leftarrow}) \pm \sigma(^{\rightarrow} _{\Rightarrow})\right).
\end{equation}
To calculate  the cross sections, we  integrate the amplitudes
squared over the $Q \bar Q$ phase space. The spin-average and
spin-dependent cross section can be written in the form
\begin{equation}
\label{sigma} \frac{d^5 \sigma(\pm)}{dQ^2 dy dx_p dt dk_\perp^2}=
\left(^{(2-2 y+y^2)} _{\hspace{3mm}(2-y)}\right)
 \frac{C(x_P,Q^2) \; N(\pm)}
{\sqrt{1-4(k_\perp^2+m_q^2)/M_X^2}}.
\end{equation}
Here $C(x_P,Q^2)$ is a normalization function which is common for
the spin average and spin dependent cross section; $N(\pm)$ is
determined by a sum of graphs integrated over the gluon momenta
$l$ and $l'$

The  contribution of the planar graphs  to the function $N(+)$ can
be approximated as
\begin{equation}\label{nn}
N^p(+) \sim \frac{\left(|\tilde B|^2+|t|/m^2 |\tilde K|^2 \right)
\left((k_\perp+r_\perp)^2+m_q^2 \right)}{\left(k_\perp^2+m_q^2
\right)^3}
\end{equation}
with
\begin{eqnarray}\label{bqq}
\tilde B \sim \int^{l_\perp^2<k_0^2}_0 \frac{d^2l_\perp
(l_\perp^2+\vec l_\perp \vec r_\perp) }
{(l_\perp^2+\lambda^2)((\vec l_\perp+\vec r_\perp)^2+\lambda^2)}
B(t,l_\perp^2,x_P,...) =  {\cal F}^g_{x_P}(x_P,t,k_0^2)\nonumber\\
\tilde K \sim \int^{l_\perp^2<k_0^2}_0 \frac{d^2l_\perp
(l_\perp^2+\vec l_\perp \vec r_\perp) }
{(l_\perp^2+\lambda^2)((\vec l_\perp+\vec r_\perp)^2+\lambda^2)}
K(t,l_\perp^2,x_P,...) =  {\cal K}^g_{x_P}(x_P,t,k_0^2),
\end{eqnarray}
where $k_0^2 \sim k_\perp^2+m_q^2$.  The details of calculations
can be found in \cite{golostr}. Connection of the two-gluon
structures with gluon the GPD \cite{golostr} is shown in
(\ref{bqq}).

The contribution of all graphs to the function $N(+)$  can be
written as
\begin{equation}\label{np}
N(+)=\left(|\tilde B|^2+|t|/m^2 |\tilde K|^2 \right)
\Pi^{(+)}(t,k_\perp^2,Q^2).
\end{equation}
The function $\Pi^{(+)}$ here has been calculated numerically.

The same analysis was carried out for the spin-dependent cross
sections. We find in the spin-dependent part the terms
proportional to the scalar production $\vec Q \vec S_\perp$ and
$\vec k_\perp \vec S_\perp$ where $S_\perp$ is a transverse
polarization of the proton
\begin{eqnarray}\label{nm}
N(-)=\sqrt{\frac{|t|}{m^2}} \left(\tilde B \tilde K^*+\tilde B^*
\tilde K\right) [ \frac{(\vec Q \vec S_\perp)}{m}
\Pi^{(-)}_Q(t,k_\perp^2,Q^2) \nonumber\\
 +\frac{(\vec k_\perp \vec
S_\perp)}{m} \Pi^{(-)}_k(t,k_\perp^2,Q^2)].
\end{eqnarray}

The spin-average cross section of the vector meson production at a
small momentum transfer is approximately proportional to the
$|\tilde B|^2$ function (\ref{np}) which is connected with the
generalized gluon distribution ${\cal F}^g$.  We use the  simple
parameterization of the GPD as a product of the form factor and
the ordinary gluon distribution
\begin{equation}
\label{b_g} \tilde B(t,x_P, \bar Q^2) =f(t) \left( x_P g(x_P,\bar
Q^2)  \right).
\end{equation}
A more general form of GPD like (\ref{f1}) can be used, but the
enhancement factor which appears in this case at small $x$ will be
canceled in the asymmetry. The form factor $F_B(t)$ in (\ref{b_g})
is chosen as an electromagnetic form factor of the proton. Such a
simple choice can be justified by that the Pomeron--proton vertex
might be similar to the photon--proton coupling
\begin{equation}
\label{fp} f(t) \sim F^{em}_p(t)=\frac{(4 m_p^2+2.8 |t|)}{(4
m_p^2+|t|)( 1+|t|/0.7GeV^2)^2}.
\end{equation}

The energy dependence of the cross sections is determined by the
Pomeron contribution to the gluon distribution function at small
$x$
\begin{equation} \label{xgx}
\label{g_x} x_P g(x_P,\bar Q^2) \sim
\frac{const}{x_P^{\alpha_p(0)-1}}.
\end{equation}
Here $\alpha_p(t)$ is a Pomeron trajectory which is chosen in the
form
\begin{equation}\label{pom}
  \alpha_p(t)=1+\epsilon+ \alpha' t
\end{equation}
with $\epsilon=0.15$ and $ \alpha'=0$. These values are in
accordance with the fit of the diffractive $J/\Psi$ production by
ZEUS \cite{zeus97}. Note that the form (\ref{xgx}) is similar to
(\ref{xgxq2}).

 We shall discuss here our prediction for the polarized diffractive $Q
\bar Q$ production determined by the $\vec k_\perp \vec S_\perp$
term in (\ref{nm}). The results for the $\vec Q_\perp \vec
S_\perp$ contribution to asymmetry can be found in \cite{golostr}.
The asymmetry is determined by the ratio of spin--dependent and
spin--average cross section (\ref{spm}) (see (\ref{nm}) and
(\ref{np})). It is approximately proportional   to the ratio of
the gluon distribution functions
\begin{equation}\label{cltqq}
 A_{LT} \sim C \, \frac{|\tilde K|}{|\tilde B|}
  \sim C \, \frac{{\cal K}^g_\zeta(\zeta)}
 {{\cal F}^g_\zeta(\zeta)}
 \;\;\;\mbox{with} \;\zeta=x_P.
\end{equation}\\[3mm]

\begin{figure}[h]
\centering \mbox{\epsfysize=80mm\epsffile{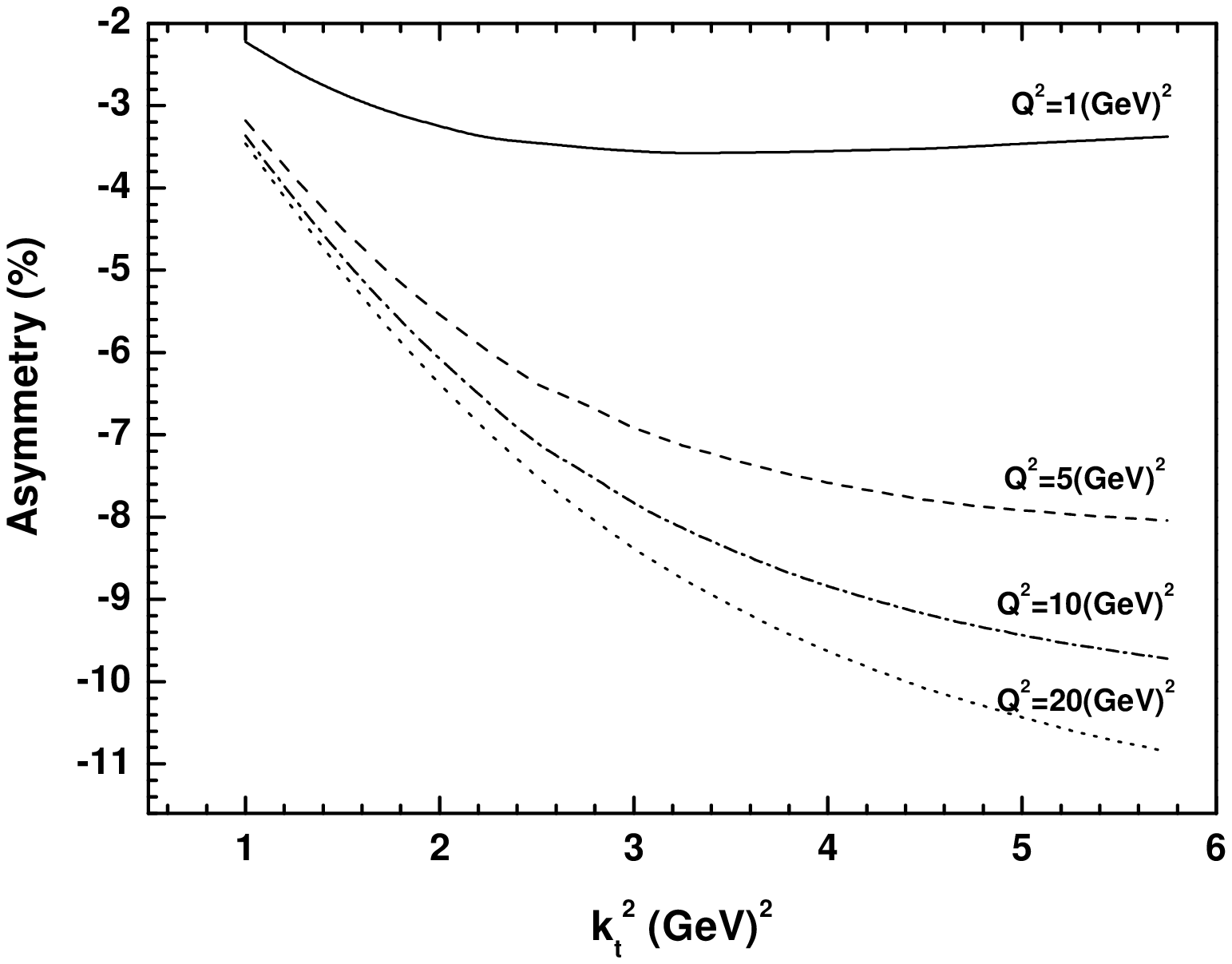}}
\caption{The $A^k_{lT}$ asymmetry in diffractive light $Q \bar Q$
production at $\sqrt{s}=50 \mbox{GeV}$ for $x_P=0.05$, $y=0.3$,
$|t|=0.3 \mbox{GeV}^2$} \label{kt_l}
\end{figure}

Thus, to estimate the value of asymmetry, we must know the ratio
of spin-dependent and  spin-average structure functions. There are
some models (see e.g. \cite{gol_mod,gol_kr}) that provide
spin-flip effects which do not vanish at high energies. These
models describe experimental data on single spin transverse
asymmetry $A_N$ \cite{krish} quite well. Predictions of other
models which lead to  spin--flip effects in the Pomeron amplitude
were discussed, e.g., in \cite{kopel}. Thus, the weak energy
dependence of spin asymmetries in exclusive reactions is not now
in contradiction with the experiment \cite{gol_mod,akch}. The
ratio of the gluon  structure functions
\begin{equation}\label{ratio}
\frac{|\tilde K|}{|\tilde B|} \sim 0.1
\end{equation}
found in \cite{gol_mod,gol_kr} for elastic scattering will be used
in our estimations of the asymmetry in diffractive hadron
production.

The term  $\vec k_\perp \vec S_\perp$ in the asymmetry is
calculated for the case when the transverse jet momentum $\vec
k_\perp$ is parallel to the target polarization $\vec S_\perp$
which gets the maximal value of asymmetry. To observe this
contribution to asymmetry, it is necessary to distinguish
experimentally the quark and antiquark jets. This can be realized
presumably by the charge of the leading particles in the jet which
should be connected in charge with the quark produced in the
photon-gluon fusion. If we do not separate events with $\vec
k_\perp$ for the quark jet,  the resulting asymmetry will be equal
to zero because the transverse momentum of the quark and antiquark
are equal and opposite in sign.

 The predicted asymmetry for light quark
production at eRHIC energies is shown in Fig. 2. The asymmetry for
heavy quark production is  approximately of the same order of
magnitude \cite{golostr}. The expected asymmetries are not small.

Possible event kinematics  in the eRHIC asymmetric system was
estimated for $p_l=5\mbox{GeV}, p_P=100\mbox{GeV}$. The energy in
lepton-proton system  $\sqrt{s} \sim 50 \mbox{GeV}$ in this case.
It was found that the final proton moves practically in the same
direction as the initial one. The jets from the final quarks have
large angles and can be detected by the eRHIC spectrometer. This
shows a possibility of studying the polarized gluon distribution
${\cal K}^g_\zeta(x)$ in future eRHIC experiment.

\section{Conclusion}
We conclude that the modified perturbative approach which
considers transverse degrees of freedom and Sudakov suppressions
in the subprocess gives a quantitative fit of the $\sigma_L$ cross
section for the $\rho$ and $\phi$ meson production. This approach
can  be used to regularize  the infrared divergencies \cite{man}
in the $\gamma^*_T\to V_L$ and $\gamma^*_T\to V_T$ amplitudes.

The analysis of $A_{LT}$ asymmetry in the diffractive $Q \bar Q$
leptoproduction at small $x$ at eRHIC energies is based on the
two-gluon spin-dependent exchange model. The spin asymmetry is
found to be proportional to the ratio of polarized gluon GPD
(\ref{cltqq}). For the $Q \bar Q$ production we have two different
terms in the $A_{lT}$ asymmetry which are proportional to $\vec
k_\perp \vec S_\perp$ and the term $\vec Q \vec S_\perp$
 (\ref{nm}). These terms in the
asymmetry have different kinematic properties and can be studied
independently.

The term $\propto \vec k_\perp \vec S_\perp$ in asymmetry has a
 coefficient $C_k$ that is predicted to be large about 0.3-0.5. This
asymmetry might be an excellent object to study transverse effects
in the proton-- gluon coupling. However, its experimental study is
not so simple. To find nonzero asymmetry in this case, it is
necessary to distinguish quark and antiquark jets and to have a
possibility of studying the azimuthal asymmetry structure. The
expected $A_{lT}$ asymmetry for the term $\propto \vec Q \vec
S_\perp$  is not small too. The predicted coefficient $C_Q$ in
this case is about 0.1-0.2 \cite{golostr}.  Note that the
asymmetry of the same order of magnitude was predicted for the
diffractive $Q \bar Q$ production in polarized proton- proton
interaction \cite{golos96}. These results for light quark
production can be used at small $x<0.1$ where the contribution of
the quark GPD is expected to be small.

Thus, our estimations show that  the  coefficient $C$ in double
spin asymmetry for longitudinally polarized leptons and
transversely polarized protons might not be small. This shows
possibility to study $A_{lT}$ asymmetry at the future eRHIC
spectrometer where the information on the polarized gluon GPD
can be obtained.\\

This work is supported  in part by the Russian Foundation for
Basic Research, Grant 03-02-16816 and by the Heisenberg-Landau
Program.

\end{document}